%% file: main.tex
\documentclass[conference]{IEEEtran} 
\usepackage{graphicx}
\usepackage{epsfig}
\usepackage{epstopdf}
\usepackage{amsmath,amsthm,amsfonts,amssymb}
\usepackage{multirow}
\usepackage{cite}
\usepackage{verbatim}
\usepackage{tikz}
\usetikzlibrary{arrows}
\usetikzlibrary{backgrounds}
\usepackage{ellipsis}
\usepackage{xcolor}
\usepackage{xifthen}
\usepackage{cancel}
\theoremstyle{definition} 

\theoremstyle{definition} 
\theoremstyle{definition}

\usetikzlibrary{shapes,arrows,intersections}
\usetikzlibrary{patterns,decorations}
\usetikzlibrary{mindmap}
\usetikzlibrary{calc}
\usepackage{ellipsis}
\usetikzlibrary{decorations.pathreplacing,decorations.markings,shapes.geometric,arrows,decorations.pathmorphing,decorations.footprints,fadings,calc,trees,mindmap,shadows,decorations.text,patterns,positioning,shapes,matrix,fit,chains}

\newcounter{mytempeqncnt}
\input{figures/BlocksStyle}
\input{figures/commands.tex}

\input{figures/HighlightingStyle}
\input{figures/LinesStyle}
\input{figures/SumNodesStyle}

\begin{document}

\title{Multi-Kernel Construction of Polar Codes}

\author{\IEEEauthorblockN{Fr\'ed\'eric Gabry, Valerio Bioglio, Ingmar Land, Jean-Claude Belfiore}
\IEEEauthorblockA{Mathematical and Algorithmic Sciences Lab\\ France Research Center, Huawei Technologies Co. Ltd.\\
Email: $\{$frederic.gabry,valerio.bioglio,ingmar.land, jean.claude.belfiore$\}$@huawei.com}} 

\maketitle

\begin{abstract}
We propose a generalized construction for binary polar codes based on mixing multiple kernels of different sizes in order to construct polar codes of block lengths that are not only powers of integers. 
This results in a multi-kernel polar code with very good performance while the encoding complexity remains low and the decoding follows the same general structure as for the original Arikan polar codes. 
The construction provides numerous practical advantages as more code lengths can be achieved without puncturing or shortening. 
We observe numerically that the error-rate performance of our construction outperforms state-of-the-art constructions using puncturing methods.
\end{abstract}

\section{Introduction}
\label{sec:intro}
\input{introduction.tex}

\section{Multi-Kernel Construction}
\label{sec:model}
\input{construction.tex}

\section{Example: Mixing $T_2$ and $T_3$ Kernels}
\label{sec:mixed_2_3}
\input{mixed_bin_ter.tex}

\section{Numerical Illustrations}
\label{sec:num}
\input{numerical_illustrations.tex}

\section{Conclusions}
\label{sec:conclusions}
\input{future_work.tex}

\bibliographystyle{IEEEbib}
\bibliography{polarbib}

\end{document}

%% file: figures/BlocksStyle.tex
\tikzset
{
	BlocksStyle/.style =
	{
		shape			= rectangle,			% shape
		rounded corners	= 0.0cm,				% radius of the rounded corner
		minimum height	= 0.7cm,				% | minimum size of the node
		minimum width	= 0.9cm,				% |
		rotate			= 0,					% angle of rotation
		scale			= 1.0,					% scaling factor
		draw			= black,				% draw the border with this color
		line width		= 0.00cm,				% thickness
		align			= center,				% text alignment
		text			= black,				% color of the fonts
		font			= \normalsize\normalfont,	% shape and dimension of the font
		inner xsep		= 0.2cm,				% min. dist. btw text and borders along x dimension
		inner ysep		= 0.2cm,				% min. dist. btw text and borders along x dimension
	}
}

\tikzset
{
	BlocksStyleb/.style =
	{
		shape			= rectangle,			% shape
		rounded corners	= 0.0cm,				% radius of the rounded corner
		minimum height	= 0.7cm,				% | minimum size of the node
		minimum width	= 0.9cm,				% |
		rotate			= 0,					% angle of rotation
		scale			= 1.0,					% scaling factor
		draw			= black,				% draw the border with this color
		line width		= 0.02cm,				% thickness
		align			= center,				% text alignment
		text			= black,				% color of the fonts
		font			= \normalsize\normalfont,	% shape and dimension of the font
		inner xsep		= 0.2cm,				% min. dist. btw text and borders along x dimension
		inner ysep		= 0.2cm,				% min. dist. btw text and borders along x dimension
	}
}

\tikzset
{
	WideBlocksStyle/.style =
	{
		BlocksStyle,
		text width		= 2.0cm,				% max. width of the text
	}
}

%% file: figures/commands.tex
\pgfmathsetseed{1}
\usetikzlibrary{shapes,positioning,arrows,calc,graphs}
\usetikzlibrary{decorations.pathreplacing,decorations.markings,shapes.geometric}
\tikzset{naming/.style={align=center,font=\small}}
\tikzset{antenna/.style={insert path={-- coordinate (ant#1) ++(0,0.25) -- +(135:0.25) + (0,0) -- +(45:0.25)}}}
\tikzset{station/.style={naming,draw,shape=dart,shape border rotate=90, minimum width=10mm, minimum height=10mm,outer sep=0pt,inner sep=3pt}}
\tikzset{mobile/.style={naming,draw,shape=rectangle,minimum width=12mm,minimum height=6mm, outer sep=0pt,inner sep=3pt}}
\tikzset{radiation/.style={{decorate,decoration={expanding waves,angle=90,segment length=4pt}}}}

%% file: figures/HighlightingStyle.tex
\tikzset
{
	HighlightingStyle/.style =
	{
		color				= blue,	
		line width			= 0.04cm,			% thickness
		arrows				= -,				% starting-ending arrows
		dotted,
	}
}	% no semicolons are here required

\tikzset
{
	HighlightingStyleD/.style =
	{
		color				= blue,	% color
		line width			= 0.04cm,			% thickness
		arrows				= -,				% starting-ending arrows
		dashed,
	}
}	% no semicolons are here required

\tikzset
{
	HighlightingStyleB/.style =
	{
		color				= black,	% color
		line width			= 0.04cm,			% thickness
		arrows				= -,				% starting-ending arrows
		solid,
	}
}

\tikzset
{
	HighlightingStyleE/.style =
	{
		color				= red,	% color
		line width			= 0.05cm,			% thickness
		arrows				= -,				% starting-ending arrows
		dashed,
	}
}	
\tikzset
{
	HighlightingStyleC/.style =
	{
		color				= black,
		fill=white,
		text=black,	% color
		line width			= 0.05cm,			% thickness
		arrows				= -,				% starting-ending arrows
 		rounded corners		= 0.0cm,			%
		solid,
	}
}

%% file: figures/LinesStyle.tex
\tikzset
{
	LinesStyle/.style =
	{
		color				= black,	% color
		line width			= 0.02cm,			% thickness
		solid,
	}
}	

\tikzset
{
	LinesStyleC/.style =
	{
		color				= black,	% color
		line width			= 0.08cm,			% thickness
		solid,
	}
}

\tikzset
{
	LinesStyleR/.style =
	{
		color				= black,	% color
		line width			= 0.08cm,			% thickness
		densely dotted,
	}
}

\tikzset
{
	LinesStyleE/.style =
	{
		color				= red,	% color
		line width			= 0.06cm,			% thickness
		solid,
	}
}

\tikzset
{
	LinesStyleb/.style =
	{
		color				= black,	% color
		line width			= 0.02cm,			% thickness
		arrows				= -latex',			% starting-ending arrows
 		shorten <			= 0.1cm,			% shorten the ending point
		solid,
	}
}

%% file: figures/SumNodesStyle.tex
\tikzset
{
	SumNodesStyle/.style =
	{
		shape			= circle,				% shape
		minimum size	= 0.1cm,				% | minimum size of the node
		rotate			= 0,					% angle of rotation
		scale			= 0.6,					% scaling factor
		draw			= black,				% draw the border with this color
		line width		= 0.02cm,				% thickness
		text			= black,				% color of the fonts
		font			= \normalsize\normalfont,	% shape and dimension of the font
	}
}

%% file: introduction.tex
Polar codes, recently introduced by Arikan in \cite{polar}, are a new class of channel codes. 
They are provably capacity-achieving over various classes of channels, and they provide excellent error rate performance for practical code lengths. 
In their original construction, polar codes are based on the polarization effect of the Kronecker powers of the binary kernel matrix
$T_2 = \begin{pmatrix}
     1 & 0 \\ 1 & 1 
\end{pmatrix}$.  
The generator matrix of a polar code is a sub-matrix of the transformation matrix $T_2^{\otimes n}$. 

In \cite{polar}, Arikan conjectured that the polarization phenomenon is not restricted to the powers of the kernel $T_2$. 
This conjecture has been proven in \cite{exp_urbanke}, where necessary and sufficient conditions are presented for kernels $T_l$ with $l>2$ to allow for the polarization effect. 
This allowed researchers to propose polar code constructions based on larger kernels \cite{kernel_presman}, both linear and non-linear \cite{non_bin_ker}. 
Moreover, generalizations of binary kernels over larger alphabets can improve the asymptotic error probability \cite{nb_polar}. 
Binary and non-binary kernels are mixed in \cite{mixed_kernel}, showing an additional improvement over homogeneous kernels construction. 
Equipped with all these techniques, it is possible to design codes based on the polarization effect of any size of the form $N = l^n$ over various alphabets.

However, many block lengths cannot be expressed in the form $N= l^n$. 
To overcome this limitation, puncturing \cite{chen_kai_punc} \cite{isit_punc} and shortening \cite{wang_liu} techniques have been proposed in the literature. 
Both shortening and puncturing techniques provide a practical way to construct polar codes of arbitrary lengths based on mother polar codes of length $N= 2^n$, albeit with some disadvantages. 
First, punctured and shortened codes are decoded on the graph of their mother codes, and therefore the decoding complexity 
can be very high with respect to the shortened code length. 
Second, puncturing and shortening may lead to a substantial loss in terms of polarization speed, and hence a worse error-rate performance. 
Finally, the lack of structure of the frozen sets and puncturing or shortening patterns generated by these methods makes them non-suitable for practical implementation. 

In this paper, we propose a generalized construction of polar codes based on mixing of kernels of different sizes over the same binary alphabet, in order to construct polar codes of any block length. 
By using kernels of different sizes in different stages, it is in fact possible to construct polar codes of block lengths that are not only powers of integers. 
This is a major difference over other mixed constructions, e.g. \cite{mixed_kernel}, designed to optimize the performance of polar codes of block length restricted to powers of integers.
With our construction, we obtain a new family of polar codes, coined \textit{multi-kernel polar codes} in the following, with error-correcting performance comparable or even better than state-of-the-art polar codes obtained via puncturing/shortening methods. 
Furthermore, the encoding complexity is similar to polar codes while the decoding follows the same general structure. 
Therefore, list decoding algorithms can be used for multi-kernel polar codes, which can as well be enhanced by the use of cyclic redundancy check (CRC) bits \cite{list_decoding}. 
In the following, we show an example of multi-kernel polar construction mixing kernels of sizes 2 and 3, which allows for code lengths $N = 2^{n_2} \cdot 3^{n_3}$ without puncturing or shortening. 

This paper is organized as follows. 
In Section \ref{sec:model}, we present our general construction, including the encoding and decoding, of multi-kernel polar codes. 
In Section \ref{sec:mixed_2_3} we describe explicitly the construction for the case of mixed binary kernels of size 2 and 3. 
In Section \ref{sec:num} we illustrate numerically the performance of the codes, and Section \ref{sec:conclusions} concludes this paper.

%% file: construction.tex
\begin{figure}[h!]
\begin{center}
\resizebox{0.55\textwidth}{!}{\begin{tikzpicture}
[
	xscale	= 1,	% to scale horizontally everything but the text
	yscale	= 1,	% to scale vertically everything but the text
]

% ------------------------------------------------------
% NODES DEFINITION
\matrix
(nMatrix)
[
	row sep		= 0.6cm,
	column sep	= 2.1cm, ampersand replacement = \&
]
{

\node (n01) {$u_0$};  \&[-4ex]\node (n02)  {};\&[-4ex]\node (n03) {};\&[-4ex]\node (n04) {};\&[-4ex]\node (n05) {};\&[-4ex]\node (n06) {};\&[-4ex]\node (n07) {};\&[-4ex]\node (n08) {$x_0 \rightarrow$ \textcolor{black}{(LLR($y_0$))}};
	\\
\node (n11) {$u_1$};\&\node(n12) {};\&\node (n13) {};\&\node (n14) {};\&\node (n15) {};\&\node (n16) {};\&\node (n17) {};\&\node (n18) {$x_1 \rightarrow$ \textcolor{black}{(LLR($y_1$))}}; \\

\node (n21) {$u_2$};\&\node(n22) {};\&\node(n23) {};\&\node (n24) {};\&\node (n25) {};\&\node (n26) {};\&\node (n27) {};\&\node (n28) {$x_2 \rightarrow$ \textcolor{black}{(LLR($y_2$))}};\\
% row 4
\node (n31) {$u_3$};\&\node(n32) {};\&\node(n33) {};\&\node(n34) {};\&\node (n35) {};\&\node (n36) {};\&\node (n37) {};\&\node (n38) {$x_3 \rightarrow$ \textcolor{black}{(LLR($y_3$))}};\\
% row 5
\node (n41) {$u_4$};\&\node(n42) {};\&\node(n43) {};\&\node(n44) {};\&\node(n45) {};\&\node (n46) {};\&\node (n47) {};\&\node (n48) {$x_4 \rightarrow$ \textcolor{black}{(LLR($y_4$))}};\&\\
% row 6
\node (n51) {$u_5$};\&\node (n52) {};\&\node(n53) {};\&\node(n54) {};\&\node(n55) {};\&\node (n56) {};\&\node (n57) {};\&\node(n58) {$x_5 \rightarrow$ \textcolor{black}{(LLR($y_5$))}};\&\\
\node (n61) {$u_6$};\&\node (n62) {};\&\node(n63) {};\&\node(n64) {};\&\node(n65) {};\&\node (n66) {};\&\node (n67) {};\&\node(n68) {$x_6 \rightarrow$ \textcolor{black}{(LLR($y_6$))}};\&\\
\node (n71) {$u_7$};\&\node (n72) {};\&\node(n73) {};\&\node(n74) {};\&\node(n75) {};\&\node (n76) {};\&\node (n77) {};\&\node(n78) {$x_7 \rightarrow$ \textcolor{black}{(LLR($y_7$))}};\&\\
};

% ------------------------------------------------------
% PATHS

%\draw[decorate,decoration={brace,amplitude=10pt},rotate=180] ($(n06)+ (0.0cm,-0.2cm)$) -- ($(n56)+ (0cm,0.2cm)$) node [pos=0.5, xshift = 0.5cm,rotate=270] {$L$ smallest metrics};

%%% -------------------------
%% auxiliary nodes

\draw [LinesStyle] (n01) -- (n08) ;

\draw [LinesStyle] (n11) -- ($(n13.west) + (-.3cm,-0cm)$) ;
\draw [LinesStyle] ($(n15.west) + (-.3cm,-0cm)$) -- ($(n13.east) + (.3cm,-0cm)$) ;
\draw [LinesStyle] ($(n17.west) + (-.3cm,-0cm)$) -- ($(n15.east) + (.3cm,-0cm)$) ;
\draw [LinesStyle] ($(n13.west) + (-.3cm,-0cm)$) --  ($(n43.east) + (.3cm,-0cm)$);
\draw [LinesStyle] ($(n15.west) + (-.3cm,-0cm)$) -- ($(n25.east) + (.3cm,-0cm)$) ;
\draw [LinesStyle] ($(n17.west) + (-.3cm,-0cm)$) -- ($(n47.east) + (.3cm,-0cm)$) ;
\draw [LinesStyle] ($(n17.east) + (.3cm,-0cm)$) -- (n18) ;

\draw [LinesStyle] (n21) -- ($(n23.west) + (-.3cm,-0cm)$) ;
\draw [LinesStyle] ($(n25.west) + (-.3cm,-0cm)$) -- ($(n23.east) + (.3cm,-0cm)$) ;
\draw [LinesStyle] ($(n27.west) + (-.3cm,-0cm)$) -- ($(n25.east) + (.3cm,-0cm)$) ;
\draw [LinesStyle] ($(n23.west) + (-.3cm,-0cm)$) --  ($(n13.east) + (.3cm,-0cm)$);
\draw [LinesStyle] ($(n25.west) + (-.3cm,-0cm)$) -- ($(n15.east) + (.3cm,-0cm)$) ;
\draw [LinesStyle] ($(n27.west) + (-.3cm,-0cm)$) -- ($(n27.east) + (.3cm,-0cm)$) ;
\draw [LinesStyle] ($(n27.east) + (.3cm,-0cm)$) -- (n28) ;

\draw [LinesStyle] (n31) -- ($(n33.west) + (-.3cm,-0cm)$) ;
\draw [LinesStyle] ($(n35.west) + (-.3cm,-0cm)$) -- ($(n33.east) + (.3cm,-0cm)$) ;
\draw [LinesStyle] ($(n37.west) + (-.3cm,-0cm)$) -- ($(n35.east) + (.3cm,-0cm)$) ;
\draw [LinesStyle] ($(n33.west) + (-.3cm,-0cm)$) --  ($(n53.east) + (.3cm,-0cm)$);
\draw [LinesStyle] ($(n35.west) + (-.3cm,-0cm)$) -- ($(n35.east) + (.3cm,-0cm)$) ;
\draw [LinesStyle] ($(n37.west) + (-.3cm,-0cm)$) -- ($(n67.east) + (.3cm,-0cm)$) ;
\draw [LinesStyle] ($(n37.east) + (.3cm,-0cm)$) -- (n38) ;

\draw [LinesStyle] (n41) -- ($(n43.west) + (-.3cm,-0cm)$) ;
\draw [LinesStyle] ($(n45.west) + (-.3cm,-0cm)$) -- ($(n43.east) + (.3cm,-0cm)$) ;
\draw [LinesStyle] ($(n47.west) + (-.3cm,-0cm)$) -- ($(n45.east) + (.3cm,-0cm)$) ;
\draw [LinesStyle] ($(n43.west) + (-.3cm,-0cm)$) --  ($(n23.east) + (.3cm,-0cm)$);
\draw [LinesStyle] ($(n45.west) + (-.3cm,-0cm)$) -- ($(n45.east) + (.3cm,-0cm)$) ;
\draw [LinesStyle] ($(n47.west) + (-.3cm,-0cm)$) -- ($(n17.east) + (.3cm,-0cm)$) ;
\draw [LinesStyle] ($(n47.east) + (.3cm,-0cm)$) -- (n48) ;

\draw [LinesStyle] (n51) -- ($(n53.west) + (-.3cm,-0cm)$) ;
\draw [LinesStyle] ($(n55.west) + (-.3cm,-0cm)$) -- ($(n53.east) + (.3cm,-0cm)$) ;
\draw [LinesStyle] ($(n57.west) + (-.3cm,-0cm)$) -- ($(n55.east) + (.3cm,-0cm)$) ;
\draw [LinesStyle] ($(n53.west) + (-.3cm,-0cm)$) --  ($(n63.east) + (.3cm,-0cm)$);
\draw [LinesStyle] ($(n55.west) + (-.3cm,-0cm)$) -- ($(n65.east) + (.3cm,-0cm)$) ;
\draw [LinesStyle] ($(n57.west) + (-.3cm,-0cm)$) -- ($(n57.east) + (.3cm,-0cm)$) ;
\draw [LinesStyle] ($(n57.east) + (.3cm,-0cm)$) -- (n58) ;

\draw [LinesStyle] (n61) -- ($(n63.west) + (-.3cm,-0cm)$) ;
\draw [LinesStyle] ($(n65.west) + (-.3cm,-0cm)$) -- ($(n63.east) + (.3cm,-0cm)$) ;
\draw [LinesStyle] ($(n67.west) + (-.3cm,-0cm)$) -- ($(n65.east) + (.3cm,-0cm)$) ;
\draw [LinesStyle] ($(n63.west) + (-.3cm,-0cm)$) --  ($(n33.east) + (.3cm,-0cm)$);
\draw [LinesStyle] ($(n65.west) + (-.3cm,-0cm)$) -- ($(n55.east) + (.3cm,-0cm)$) ;
\draw [LinesStyle] ($(n67.west) + (-.3cm,-0cm)$) -- ($(n37.east) + (.3cm,-0cm)$) ;
\draw [LinesStyle] ($(n67.east) + (.3cm,-0cm)$) -- (n68) ;

\draw [LinesStyle] (n71) -- (n78) ;

\node [coordinate, xshift = -0.3cm, yshift =  0.3cm] (nAux0) at (n03.north west) {};
\node [coordinate, xshift =  0.3cm, yshift =  -0.3cm] (nAux00) at (n73.south east) {};
\draw [HighlightingStyle]  (nAux0) -| (nAux00) -|  (nAux0)
node [below, pos = 0.21] {$P_3$};

\node [coordinate, xshift = -0.3cm, yshift =  0.3cm] (nAux1) at (n05.north west) {};
\node [coordinate, xshift =  0.3cm, yshift =  -0.3cm] (nAux2) at (n75.south east) {};
\draw [HighlightingStyle]  (nAux1) -| (nAux2) -|  (nAux1)
node [below, pos = 0.21] {$P_2$};

\node [coordinate, xshift = -0.3cm, yshift =  0.3cm] (nAux3) at (n07.north west) {};
\node [coordinate, xshift =  0.3cm, yshift =  -0.3cm] (nAux4) at (n77.south east) {};
\draw [HighlightingStyle] (nAux3) -| (nAux4) -|  (nAux3)
node (Pnode) [below, pos = 0.21] {$P_1$};

\node [coordinate, xshift = -0.3cm, yshift =  0.2cm] (nAux5) at (n02.north west) {};
\node [coordinate, xshift = 0.2cm, yshift =  -0.2cm] (nAux6) at (n12.south east) {};
\draw [HighlightingStyleB,fill=white] (nAux5) -| (nAux6) -|  (nAux5)
node (Pnode1) [xshift=0.4cm, yshift =-.9cm] {$T_2$};

\node [coordinate, xshift = -0.3cm, yshift =  0.2cm] (nAux7) at (n22.north west) {};
\node [coordinate, xshift = 0.2cm, yshift =  -0.2cm] (nAux8) at (n32.south east) {};
\draw [HighlightingStyleB,fill=white] (nAux7) -| (nAux8) -|  (nAux7)
node (Pnode1) [xshift=0.4cm, yshift =-.9cm] {$T_2$};

\node [coordinate, xshift = -0.3cm, yshift =  0.2cm] (nAux9) at (n42.north west) {};
\node [coordinate, xshift = 0.2cm, yshift =  -0.2cm] (nAux10) at (n52.south east) {};
\draw [HighlightingStyleB,fill=white] (nAux9) -| (nAux10) -|  (nAux9)
node (Pnode1) [xshift=0.4cm, yshift =-.9cm] {$T_2$};

\node [coordinate, xshift = -0.3cm, yshift =  0.2cm] (nAux11) at (n62.north west) {};
\node [coordinate, xshift = 0.2cm, yshift =  -0.2cm] (nAux12) at (n72.south east) {};
\draw [HighlightingStyleB,fill=white] (nAux11) -| (nAux12) -|  (nAux11)
node (Pnode1) [xshift=0.4cm, yshift =-.9cm] {$T_2$};

\node [coordinate, xshift = -0.3cm, yshift =  0.2cm] (nAux5b) at (n04.north west) {};
\node [coordinate, xshift = 0.2cm, yshift =  -0.2cm] (nAux6b) at (n14.south east) {};
\draw [HighlightingStyleB,fill=white] (nAux5b) -| (nAux6b) -|  (nAux5b)
node (Pnode1) [xshift=0.4cm, yshift =-.9cm] {$T_2$};

\node [coordinate, xshift = -0.3cm, yshift =  0.2cm] (nAux7b) at (n24.north west) {};
\node [coordinate, xshift = 0.2cm, yshift =  -0.2cm] (nAux8b) at (n34.south east) {};
\draw [HighlightingStyleB,fill=white] (nAux7b) -| (nAux8b) -|  (nAux7b)
node (Pnode1) [xshift=0.4cm, yshift =-.9cm] {$T_2$};

\node [coordinate, xshift = -0.3cm, yshift =  0.2cm] (nAux9b) at (n44.north west) {};
\node [coordinate, xshift = 0.2cm, yshift =  -0.2cm] (nAux10b) at (n54.south east) {};
\draw [HighlightingStyleB,fill=white] (nAux9b) -| (nAux10b) -|  (nAux9b)
node (Pnode1) [xshift=0.4cm, yshift =-.9cm] {$T_2$};

\node [coordinate, xshift = -0.3cm, yshift =  0.2cm] (nAux11b) at (n64.north west) {};
\node [coordinate, xshift = 0.2cm, yshift =  -0.2cm] (nAux12b) at (n74.south east) {};
\draw [HighlightingStyleB,fill=white] (nAux11b) -| (nAux12b) -|  (nAux11b)
node (Pnode1) [xshift=0.4cm, yshift =-.9cm] {$T_2$};

\node [coordinate, xshift = -0.3cm, yshift =  0.2cm] (nAux5c) at (n06.north west) {};
\node [coordinate, xshift = 0.2cm, yshift =  -0.2cm] (nAux6c) at (n16.south east) {};
\draw [HighlightingStyleB,fill=white] (nAux5c) -| (nAux6c) -|  (nAux5c)
node (Pnode1) [xshift=0.4cm, yshift =-.9cm] {$T_2$};

\node [coordinate, xshift = -0.3cm, yshift =  0.2cm] (nAux7c) at (n26.north west) {};
\node [coordinate, xshift = 0.2cm, yshift =  -0.2cm] (nAux8c) at (n36.south east) {};
\draw [HighlightingStyleB,fill=white] (nAux7c) -| (nAux8c) -|  (nAux7c)
node (Pnode1) [xshift=0.4cm, yshift =-.9cm] {$T_2$};

\node [coordinate, xshift = -0.3cm, yshift =  0.2cm] (nAux9c) at (n46.north west) {};
\node [coordinate, xshift = 0.2cm, yshift =  -0.2cm] (nAux10c) at (n56.south east) {};
\draw [HighlightingStyleB,fill=white] (nAux9c) -| (nAux10c) -|  (nAux9c)
node (Pnode1) [xshift=0.4cm, yshift =-.9cm] {$T_2$};

\node [coordinate, xshift = -0.3cm, yshift =  0.2cm] (nAux11c) at (n66.north west) {};
\node [coordinate, xshift = 0.2cm, yshift =  -0.2cm] (nAux12c) at (n76.south east) {};
\draw [HighlightingStyleB,fill=white] (nAux11c) -| (nAux12c) -|  (nAux11c)
node (Pnode1) [xshift=0.4cm, yshift =-.9cm] {$T_2$};

\node[left of = Pnode,node distance = 1.8cm] {Stage 1};
\node[left of = Pnode,node distance = 5.3cm] {Stage 2};
\node[left of = Pnode,node distance = 8.5cm] {Stage 3};
\end{tikzpicture}}
\caption{Tanner graph of the binary-kernel polar code with $N=8$ and $G_8= T_2 \otimes T_2 \otimes T_2$.}
\label{fig:polar_8}
\end{center}
\end{figure}
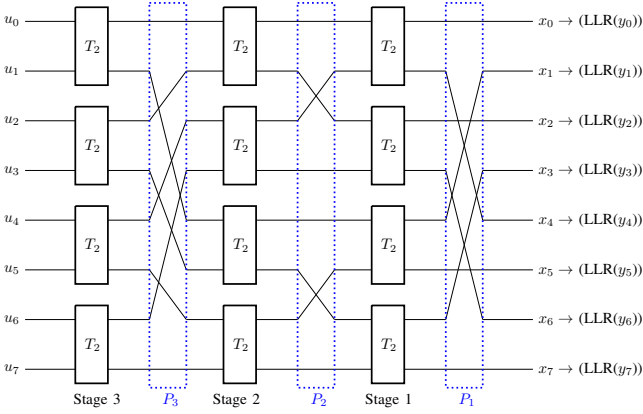

In this section, we describe the proposed multi-kernel polar codes construction for a code of length $N = n_1 \cdot \dots \cdot n_s$, with $n_i$ not being necessarily distinct prime numbers. 
Each integer $n_i$ corresponds to a binary kernel $T_{n_i}$ of size $n_i$, i.e., a squared $n_i \times n_i$ binary matrix \cite{kernel_presman}. 
The transformation matrix of such a multi-kernel polar code is defined as $G_N \triangleq T_{n_1} \otimes \cdots \otimes T_{n_s}$. 
Note that a different ordering of the kernels would drive to a different transformation matrix, as the Kronecker product is not commutative. 

\subsection{Tanner Graph Construction}

The Tanner graph of multi-kernel polar codes is a generalization of the Tanner graph of polar codes, shown in Figure~\ref{fig:polar_8}. 
This graph has $s$ stages, where $s$ is the number of kernels used to build $G_N$. 
Each stage is constituted by $B_i = N/n_i$ boxes, where each box is a $n_i \times n_i$ block depicting a $T_{n_i}$ kernel. 
As a consequence, stage 1, i.e., the rightmost stage in the graph, is constituted of $B_1$ $T_{n_1}$-boxes and so on, until the last stage $s$ constituted of $B_s$ $T_{n_s}$-boxes.

The connections between stages are generated as follows. 
We call $P_i$ the permutation connecting the outputs of boxes of stage $i-1$ to the boxes of stage $i$. We call $P_1$ the permutation connecting the encoded bits to the boxes of stage 1. 
For polar codes, $P_1$ is given by the bit-reversal permutation \cite{polar}. 
On the other hand, for multi-kernel polar codes, $P_1$ is the inverse of the product of the other permutations, i.e., $P_1 = (P_2 \cdot \dots \cdot P_s)^{-1}$. 
The other permutations are based on a basic permutation, that we call $Q_i$-canonical permutation. 
If $N_i = \prod_{j=1}^{i-1} n_j$ is the partial product of the kernel sizes at stage $i$, then $Q_i$ is a permutation of $N_{i+1}$ elements defined as in Equation \eqref{eq:Q_can} at the top of next page.
Finally, $P_i = (Q_i | \: Q_i+N_{i+1} | \: Q_i+2N_{i+1} | \: \dots | \: Q_i+(N/N_{i+1}-1)N_{i+1})$. 
Note that for the last stage $P_s = Q_s$. 
An example of this construction is showed in Figure~\ref{fig:G_6}. 

\begin{figure*}[!t]
\normalsize
\setcounter{mytempeqncnt}{\value{equation}}

\setcounter{equation}{0}
\begin{equation}
\label{eq:Q_can}
Q_i = 
\begin{pmatrix}
  1 & 2     & \dots & N_i          & N_i+1 & N_i+2 & \dots & (n_i-1)N_i+1 & \dots & N_{i+1} \\%n_i N_i \\
  1 & n_i+1 & \dots & (N_i-1)n_i+1 & 2     & n_i+2 & \dots & n_i          & \dots & N_{i+1} %n_i N_i
\end{pmatrix}.
\end{equation}
\setcounter{equation}{\value{mytempeqncnt}}
\hrulefill
\vspace*{4pt}
\end{figure*}

In order to clarify the description of the graph construction, let us construct the graph in Figure \ref{fig:polar_8}, i.e., for the $G_8=T_2 \otimes T_2 \otimes T_2$ transformation matrix, using the proposed algorithm. 
We first draw 3 stages, each constituted of $N/2 = 4$ $T_2$ boxes. 
To construct the edges between Stage 1 and Stage 2, we calculate the canonical permutation $Q_2 = \begin{pmatrix}
  1 & 2 & 3 & 4 \\
  1 & 3 & 2 & 4
\end{pmatrix}$.
Given the canonical permutation, we can calculate $P_2 = (Q_2 | \: Q_2+N_3) = 
\begin{pmatrix}
  1 & 2 & 3 & 4 & 5 & 6 & 7 & 8 \\
  1 & 3 & 2 & 4 & 5 & 7 & 6 & 8
\end{pmatrix}$. 
We observe indeed, in Figure~\ref{fig:polar_8}, that bit 1 of stage 1 is connected to bit 1 of stage 2, bit 2 is connected to bit 3, etc. 
To construct the edges between Stage 2 and Stage 3, we recall that $P_3 = Q_3$, and hence $P_3 = \begin{pmatrix}
  1 & 2 & 3 & 4 & 5 & 6 & 7 & 8 \\
  1 & 3 & 5 & 7 & 2 & 4 & 6 & 8
\end{pmatrix}$. 
Finally, $P_1 = (P_2 \cdot P_3)^{-1}= \begin{pmatrix}
  1 & 2 & 3 & 4 & 5 & 6 & 7 & 8 \\
  1 & 5 & 3 & 7 & 2 & 6 & 4 & 8
\end{pmatrix}$. 

\subsection{Encoding of Multi-Kernel Polar Codes}
As for polar codes, a multi-kernel polar code of length $N$ and dimension $K$ is completely defined by a transformation matrix $G_N=T_{n_1} \otimes \cdots \otimes T_{n_s}$ and a frozen set $\mathcal{F}$. 
We recall that, in contrast to polar codes, the order of the factors of the Kronecker product is important, since different orderings result in different transformation matrices, with different polarization behaviors and hence different frozen sets. 
Given a kernel order, the reliabilities of the bits can be calculated for a target SNR through a Monte-Carlo method or the density evolution algorithm \cite{DE_mori} on the resulting transformation matrix. 
We select the order of the kernels summing the reliabilities of the $K$ best bits of each ordering, keeping the kernel order resulting in the largest sum.   
When the ordering of the kernels is decided, along with the corresponding transformation matrix of the code, the frozen set $\mathcal{F}$ is given by the $N-K$ less reliable bits. 
To simplify the notation, in the following we assume the kernels in the Kronecker product to be already ordered. 

Hereafter, the $K$ information bits are stored in the length-$N$ message $u$ according to the frozen set, i.e., they are stored in the positions not belonging to $\mathcal{F}$, while the remaining bits are filled with zeros. 
Finally, a codeword $x$ of length $N$ is obtained as $x=u∙G_N$. 
We notice that the encoding may be performed on the Tanner graph of the multi-kernel polar code. 
In Section \ref{sec:mixed_2_3} we will give an example of this construction for the case $N=6$.

\subsection{Decoding of Multi-Kernel Polar Codes}
The decoding of multi-kernel polar codes is performed through successive cancellation (SC) decoding on the Tanner graph of the code, similarly to polar codes. 
In general, the log-likelihood ratios (LLRs) are passed along the Tanner graph from the right to the left, while the hard decisions on the decoded bits are passed from the left to the right. 
Assuming that $G_N=T_{n_1} \otimes \cdots \otimes T_{n_s}$, the LLRs go through $B_1$  $T_{n_1}$-boxes of size $n_1\times n_1$ on the first stage, until the $B_s$ $ T_{n_s}$-boxes of size $n_s$-by-$n_s$ on the last stage. 
We call LLR$(j,i)$ and $u(j,i)$ the values taken by the LLR and the hard decision of bit $i$ at stage $j$ of the Tanner graph respectively.  
The update of the LLRs is then done according to update functions corresponding to the kernel used at a given stage. 
The description of this section will be clarified in Section~\ref{sec:mixed_2_3} with a detailed description of the update functions for LLRs and hard decisions for kernels $T_2$ and $T_3$.

SC decoding operates as follows. 
Initially, the LLRs of the coded bits $x_i$ based on the received vector $\bf{y}$ are calculated at the receiver. 
The received signal is decoded bit-by-bit using LLR propagation through the graph to retrieve the transmitted message $\mathbf{u}=[u_0,\cdots,u_i,\cdots,u_{N-1}]$.
For every bit $u_i$, the nature of its position $i$ is initially checked. 
If $u_i$ is a frozen bit, it is decoded as $\hat{u}_i=0$, and the decoder moves on to the next bit.
If $u_i$ is an information bit, its LLR is recursively calculated to make a hard decision, starting by LLR$(s,i)$. 
In general, the calculation of LLR$(j,l)$ is done using the LLR$(j-1,\cdot)$ and the hard decisions $u(j,\cdot)$ that are connected to the $T_{n_j}$ box outputting LLR$(j,l)$. 
The value of LLR$(j,l)$ is calculated according to the update rules corresponding to the $T_{n_{j}}$ kernel. 
These update rules are given in Section~\ref{sec:mixed_2_3} for the case of kernels $T_2$ and $T_3$ used in the numerical illustration of the code construction.
Using such a recursive procedure, the algorithm will arrive to the calculations of LLR$(1,\cdot)$ values in the graph, which are obtained using the channel LLRs and the updating rules corresponding to $T_{n_1}$. 
Hence all the LLRs needed for the computation of LLR$(s,i)$ are eventually calculated, and LLR$(s,i)$ is computed. 
Finally, the bit is decoded as $\hat{u}_i$ given by the hard decision corresponding to the sign of LLR$(s,i)$.

%% file: mixed_bin_ter.tex
\begin{figure}[h!]
\begin{center}
\resizebox{0.55\textwidth}{!}{\begin{tikzpicture}
[
	xscale	= 1,	% to scale horizontally everything but the text
	yscale	= 1,	% to scale vertically everything but the text
]

% ------------------------------------------------------
% NODES DEFINITION
\matrix
(nMatrix)
[
	row sep		= 0.6cm,
	column sep	= 2.1cm, ampersand replacement=\&
]
{

\node (n01) {$u_0$}; \&\node (n02)  {};\& [-4ex]\node (n03) {};\&\node (n04) {};\& [-4ex]\node (n05) {};\&\node (n06) {$x_0$};
	\\
\node (n11) {$u_1$};\&\node(n12) {};\&\node (n13) {};\&\node (n14) {};\&\node (n15) {};\&\node (n16) {$x_1$};\& \\

\node (n21) {$u_2$};\&\node(n22) {};\&\node(n23) {};\&\node (n24) {};\&\node (n25) {};\&\node (n26) {$x_2$};\\
% row 4
\node (n31) {$u_3$};\&\node(n32) {};\&\node(n33) {};\&\node(n34) {};\&\node (n35) {};\&\node (n36) {$x_3$};\\
% row 5
\node (n41) {$u_4$};\&\node(n42) {};\&\node(n43) {};\&\node(n44) {};\&\node(n45) {};\&\node (n46) {$x_4$};\&\\
% row 6
\node (n51) {$u_5$};\&\node (n52) {};\&\node(n53) {};\&\node(n54) {};\&\node(n55) {};\&\node(n56) {$x_5$};\&\\
};

% ------------------------------------------------------
% PATHS

%\draw[decorate,decoration={brace,amplitude=10pt},rotate=180] ($(n06)+ (0.0cm,-0.2cm)$) -- ($(n56)+ (0cm,0.2cm)$) node [pos=0.5, xshift = 0.5cm,rotate=270] {$L$ smallest metrics};

%%% -------------------------
%% auxiliary nodes

\draw [LinesStyle] (n01) -- (n06) ;

\draw [LinesStyle] (n11) -- ($(n13.west) + (-.3cm,-0cm)$) ;
\draw [LinesStyle] ($(n13.west) + (-.3cm,-0cm)$) --  ($(n23.east) + (.3cm,-0cm)$);
%\draw [LinesStyle] (n12.east) -- (n24.west) ;
%\draw [LinesStyle] ($(b4.north east) + (0cm,-.3cm)$) -- (n16.west) ;
\draw [LinesStyle] ($(n15.west) + (-.3cm,-0cm)$) -- ($(n13.east) + (.3cm,-0cm)$) ;
\draw [LinesStyle] ($(n15.west) + (-.3cm,-0cm)$) -- ($(n35.east) + (.3cm,-0cm)$) ;
\draw [LinesStyle] ($(n15.east) + (.3cm,-0cm)$) -- (n16) ;
%
%
%\draw [LinesStyle] (n21) -- (n22.east) ;
\draw [LinesStyle] (n21) -- ($(n23.west) + (-.3cm,-0cm)$) ;
\draw [LinesStyle] ($(n23.west) + (-.3cm,-0cm)$) --  ($(n43.east) + (.3cm,-0cm)$);
\draw [LinesStyle] ($(n25.west) + (-.3cm,-0cm)$) -- ($(n23.east) + (.3cm,-0cm)$) ;
\draw [LinesStyle] ($(n25.west) + (-.3cm,-0cm)$) -- ($(n15.east) + (.3cm,-0cm)$) ;
\draw [LinesStyle] ($(n25.east) + (.3cm,-0cm)$) -- (n26) ;
%\draw [LinesStyle] ($(n22.east) + (0cm,+.18cm)$) -- ($(n44.west) + (0cm,-.33cm)$) ;
%\draw [LinesStyle] ($(b5.north east) + (0cm,-.3cm)$) -- (n26.west) ;
%
%\draw [LinesStyle] (n31) -- (n32.east) ;
\draw [LinesStyle] (n31) -- ($(n33.west) + (-.3cm,-0cm)$) ;
\draw [LinesStyle] ($(n33.west) + (-.3cm,-0cm)$) --  ($(n13.east) + (.3cm,-0cm)$);
\draw [LinesStyle] ($(n35.west) + (-.3cm,-0cm)$) -- ($(n33.east) + (.3cm,-0cm)$) ;
\draw [LinesStyle] ($(n35.west) + (-.3cm,-0cm)$) -- ($(n45.east) + (.3cm,-0cm)$) ;
\draw [LinesStyle] ($(n35.east) + (.3cm,-0cm)$) -- (n36) ;
%\draw [LinesStyle] ($(n32.east) + (0cm,-.28cm)$) -- ($(n14.west) + (0cm,+.33cm)$) ;
%\draw [LinesStyle] ($(b3.south east) + (0cm,.3cm)$) -- (n36.west) ;
%
%\draw [LinesStyle] (n41) -- (n42.east) ;
\draw [LinesStyle] (n41) -- ($(n43.west) + (-.3cm,-0cm)$) ;
\draw [LinesStyle] ($(n43.west) + (-.3cm,-0cm)$) --  ($(n33.east) + (.3cm,-0cm)$);
\draw [LinesStyle] ($(n45.west) + (-.3cm,-0cm)$) -- ($(n43.east) + (.3cm,-0cm)$) ;
\draw [LinesStyle] ($(n45.west) + (-.3cm,-0cm)$) -- ($(n25.east) + (.3cm,-0cm)$) ;
\draw [LinesStyle] ($(n45.east) + (.3cm,-0cm)$) -- (n46) ;
%\draw [LinesStyle] (n42.east) -- ($(n34.west) + (0cm,+.15cm)$) ;
%\draw [LinesStyle] ($(b4.south east) + (0cm,.3cm)$) -- (n46.west) ;

\draw [LinesStyle] (n51) -- (n56) ;

\node [coordinate, xshift = -0.3cm, yshift =  0.3cm] (nAux1) at (n03.north west) {};
\node [coordinate, xshift =  0.3cm, yshift =  -0.3cm] (nAux2) at (n53.south east) {};
\draw [HighlightingStyle]  (nAux1) -| (nAux2) -|  (nAux1)
node [below, pos = 0.21] {$P_2$};

\node [coordinate, xshift = -0.3cm, yshift =  0.3cm] (nAux3) at (n05.north west) {};
\node [coordinate, xshift =  0.3cm, yshift =  -0.3cm] (nAux4) at (n55.south east) {};
\draw [HighlightingStyle] (nAux3) -| (nAux4) -|  (nAux3)
node (Pnode) [below, pos = 0.21] {$P_1$};

\node [coordinate, xshift = -0.3cm, yshift =  0.2cm] (nAux5) at (n02.north west) {};
\node [coordinate, xshift = 0.2cm, yshift =  -0.2cm] (nAux6) at (n22.south east) {};
\draw [HighlightingStyleB,fill=white] (nAux5) -| (nAux6) -|  (nAux5)
node (Pnode1) [xshift=0.4cm, yshift =-1.4cm] {$T_3$};

\node [coordinate, xshift = -0.3cm, yshift =  0.2cm] (nAux7) at (n32.north west) {};
\node [coordinate, xshift = 0.2cm, yshift =  -0.2cm] (nAux8) at (n52.south east) {};
\draw [HighlightingStyleB,fill=white] (nAux7) -| (nAux8) -|  (nAux7)
node (Pnode1) [xshift=0.4cm, yshift =-1.4cm] {$T_3$};

\node [coordinate, xshift = -0.3cm, yshift =  0.2cm] (nAux9) at (n04.north west) {};
\node [coordinate, xshift = 0.2cm, yshift =  -0.2cm] (nAux10) at (n14.south east) {};
\draw [HighlightingStyleB,fill=white] (nAux9) -| (nAux10) -|  (nAux9)
node (Pnode1) [xshift=0.4cm, yshift =-.9cm] {$T_2$};

\node [coordinate, xshift = -0.3cm, yshift =  0.2cm] (nAux11) at (n24.north west) {};
\node [coordinate, xshift = 0.2cm, yshift =  -0.2cm] (nAux12) at (n34.south east) {};
\draw [HighlightingStyleB,fill=white] (nAux11) -| (nAux12) -|  (nAux11)
node (Pnode1) [xshift=0.4cm, yshift =-.9cm] {$T_2$};

\node [coordinate, xshift = -0.3cm, yshift =  0.2cm] (nAux13) at (n44.north west) {};
\node [coordinate, xshift = 0.2cm, yshift =  -0.2cm] (nAux14) at (n54.south east) {};
\draw [HighlightingStyleB,fill=white] (nAux13) -| (nAux14) -|  (nAux13)
node (Pnode1) [xshift=0.4cm, yshift =-0.9cm] {$T_2$};

\node[left of = n02,yshift=0.18cm,xshift=-0.05cm] (LLR20) {\footnotesize{LLR$(2,0)$}};
\node[left of = n12,yshift=0.18cm,xshift=-0.05cm] (LLR21) {\footnotesize{LLR$(2,1)$}};
\node[left of = n22,yshift=0.18cm,xshift=-0.05cm] (LLR22) {\footnotesize{LLR$(2,2)$}};
\node[left of = n32,yshift=0.18cm,xshift=-0.05cm] (LLR23) {\footnotesize{LLR$(2,3)$}};
\node[left of = n42,yshift=0.18cm,xshift=-0.05cm] (LLR24) {\footnotesize{LLR$(2,4)$}};
\node[left of = n52,yshift=0.18cm,xshift=-0.05cm] (LLR25) {\footnotesize{LLR$(2,5)$}};

\node[left of = n04,yshift=0.18cm,xshift=-0.05cm] (LLR10) {\footnotesize{LLR$(1,0)$}};
\node[left of = n14,yshift=0.18cm,xshift=-0.05cm] (LLR11) {\footnotesize{LLR$(1,1)$}};
\node[left of = n24,yshift=0.18cm,xshift=-0.05cm] (LLR12) {\footnotesize{LLR$(1,2)$}};
\node[left of = n34,yshift=0.18cm,xshift=-0.05cm] (LLR13) {\footnotesize{LLR$(1,3)$}};
\node[left of = n44,yshift=0.18cm,xshift=-0.05cm] (LLR14) {\footnotesize{LLR$(1,4)$}};
\node[left of = n54,yshift=0.18cm,xshift=-0.05cm] (LLR15) {\footnotesize{LLR$(1,5)$}};

\node[left of = n06,yshift=0.18cm,xshift=-0.05cm] (LLR00) {\footnotesize{LLR$(0,0)$}};
\node[left of = n16,yshift=0.18cm,xshift=-0.05cm] (LLR01) {\footnotesize{LLR$(0,1)$}};
\node[left of = n26,yshift=0.18cm,xshift=-0.05cm] (LLR02) {\footnotesize{LLR$(0,2)$}};
\node[left of = n36,yshift=0.18cm,xshift=-0.05cm] (LLR03) {\footnotesize{LLR$(0,3)$}};
\node[left of = n46,yshift=0.18cm,xshift=-0.05cm] (LLR04) {\footnotesize{LLR$(0,4)$}};
\node[left of = n56,yshift=0.18cm,xshift=-0.05cm] (LLR05) {\footnotesize{LLR$(0,5)$}};

\node[left of = Pnode,node distance = 1.7cm] {Stage 1};
\node[left of = Pnode,node distance = 5.8cm] {Stage 2};
\end{tikzpicture}}
\caption{Tanner graph of the multi-kernel polar code with $N=6$ and $G_6=T_2 \otimes T_3$.}
\label{fig:G_6}
\end{center}
\end{figure}
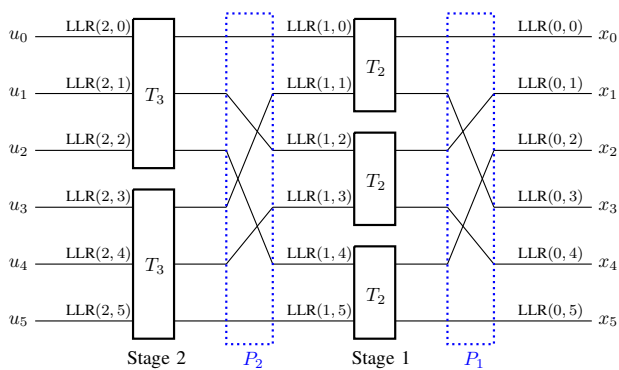

In this section, we illustrate the general code construction through a simple example using kernels $T_2$ and $T_3$, which allow us to construct codes for any length expressed as $N= 2^{l_2} \cdot 3^{l_3}$. 
In particular, we will illustrate the encoding and decoding of the proposed codes using a $N=6$ code with transformation matrix $G_6=T_2 \otimes T_3$, see Figure \ref{fig:G_6}. 
We recall that the code construction is more general, and multi-kernel polar codes mixing binary kernels of any size can be designed following the construction described in Section~\ref{sec:model}. 
We initially describe the updating rules for the kernels used in our implementation. 
Even if the update rules for $T_2$ are well-known as the canonical component of original polar codes, the update rules for $T_3$ kernel are mostly unknown in the literature. 
Then, we show how to construct the Tanner graph of multi-kernel polar codes based on binary kernels of sizes 2 and 3. 
Finally, we show how the SC decoding procedure described previously is applied to the Tanner graph of the code.

\subsection{Decoding rules and functions for $T_2$ kernels}
In this section we remind the well-known decoding rules for the Arikan polar codes' kernel \cite{polar}. The fundamental $T_2$ block can be depicted as 
    \begin{center}
		\begin{tikzpicture}[scale=0.2, line width=0.8pt]
		  \draw   (0,0) rectangle (4,4) node[midway]{$T_2$};
		  \draw[] (-2,3) node[left] {$u_0 , \lambda_0$} -- (0,3) ;
		  \draw[] (-2,1) node[left] {$u_1 , \lambda_1$} -- (0,1) ;
		  \draw[] (4,3)                     -- (6,3) node[right] {$x_0 , l_0$} ;
		  \draw[] (4,1)                     -- (6,1) node[right] {$x_1 , l_1$} ;
		\end{tikzpicture}
    \end{center}
where $( u_0 \; u_1 ) \cdot T_2 = ( x_0 \; x_1 )$, with  $T_2 = \begin{pmatrix} 1 & 0 \\ 1 & 1 \end{pmatrix}$, and where $\lambda_i$ and $l_i$ denote the LLRs and $u_i$ and $x_i$ denote the hard decisions on the bits.
This corresponds to the hard-decision update rules 
    \begin{align*}
     x_0 &= u_0 \oplus u_1,  \\
     x_1 &= u_1. 
   \end{align*}
The inverse of the update rules are $u_0 = x_0 \oplus x_1$ and $u_1 = x_1 = u_0 \oplus x_0$, corresponding to the message update equations
    \begin{align*}
      \lambda_0 &=  l_0 \boxplus l_1,  \\
      \lambda_1 &= (-1)^{u_0} \cdot l_0 + l_1,
    \end{align*}
where $ a \boxplus b \triangleq 2 \tanh^{-1} \bigl( \tanh\frac{a}{2} \cdot \tanh\frac{b}{2} \bigr)  \simeq sign(a) \cdot sign(b) \cdot \min(|a|,|b|) $. 

\subsection{Decoding rules and functions for $T_3$ kernels}
In this section we describe the decoding rules for the $T_3$ kernel used in our design. 
A study of the design of kernels of size 3 can be found in \cite{T_3_study}. 
The fundamental $T_3$ block can be depicted as follows
    \begin{center}
		\begin{tikzpicture}[scale=0.2, line width=0.8pt]
		  \draw   (0,0) rectangle (4,6) node[midway]{$T_3$};
		  \draw[] (-2,5) node[left] {$u_0 , \lambda_0$} -- (0,5) ;
		  \draw[] (-2,3) node[left] {$u_1 , \lambda_1$} -- (0,3) ;
		  \draw[] (-2,1) node[left] {$u_2 , \lambda_2$} -- (0,1) ;
		  \draw[] (4,5)                     -- (6,5) node[right] {$x_0 , l_0$} ;
		  \draw[] (4,3)                     -- (6,3) node[right] {$x_1 , l_1$} ;
		  \draw[] (4,1)                     -- (6,1) node[right] {$x_2 , l_2$} ;
		\end{tikzpicture}
    \end{center}
where $( u_0 \; u_1 \; u_2 ) \cdot T_3 = ( x_0 \; x_1 \; x_2 )$, with $T_3 \triangleq \begin{pmatrix} 1 & 1 & 1 \\ 1 & 0 & 1 \\ 0 & 1 & 1 \end{pmatrix}$. 
Consequently, the hard-decisions update rules are 
 \begin{align*}
      x_0 &= u_0 \oplus u_1, \\
      x_1 &= u_0 \oplus u_2, \\
      x_2 &= u_0 \oplus u_1 \oplus u_2.
    \end{align*}
Inverting these equations, we get $u_0 = x_0 \oplus x_1 \oplus x_2$, $u_1 = u_0 \oplus x_0 = x_1 \oplus x_2$ and $u_2 = u_0 \oplus x_1 = u_0 \oplus u_1 \oplus x_2$, from which we get the message update equations 
    \begin{align*}
      \lambda_0 &= l_0 \boxplus l_1 \boxplus l_2,  \\
      \lambda_1 &=  (-1)^{u_0} \cdot l_0 + l_1 \boxplus l_2,  \\
      \lambda_2 &=  (-1)^{u_0} \cdot l_1 + (-1)^{u_0 \oplus u_1} \cdot l_2.
    \end{align*} 

\subsection{Construction Example for $G_6=T_2 \otimes T_3$}
The Tanner graph of the length-6 code obtained from the $G_6=T_2 \otimes T_3$ transformation matrix is shown in Figure~\ref{fig:G_6}. 
Let $u=[u_0,\dots,u_5]$ be the bits to be encoded, $K$ of which are information bits and $6-K$ are frozen bits (according to the frozen set $\mathcal{F}$). 
The codeword $x=[x_0,\dots,x_5]$ is then constructed as $x = u \cdot G_6$, or following the hard-decision update rules of the Tanner graph. 

The decoding starts with  $u_0$. 
If $u_0$ is a frozen bit, its value $\hat{u}_0$ is set to 0 and the decoding continues with $u_1$. 
Otherwise, LLR$(2,0)$ is calculated to make a hard decision on the value of $u_0$. 
According to the Tanner graph and the decoding rules of the $T_3$ kernel, $\text{LLR}(2,0) = \text{LLR}(1,0) \boxplus \text{LLR}(1,2) \boxplus \text{LLR}(1,4)$. 
The values of these intermediate LLRs have to be calculated. 
According to the update rules of kernel $T_2$ and the Tanner graph of the multi-kernel polar code, $\text{LLR}(1,0) = \text{LLR}(0,0) \boxplus \text{LLR}(0,3)$, while $\text{LLR}(1,2) = \text{LLR}(0,1) \boxplus \text{LLR}(0,4)$ and $\text{LLR}(1,4) = \text{LLR}(0,2) \boxplus \text{LLR}(0,5)$. 
Since LLRs at stage 0 are the LLRs of the received signal, the recursion stops and LLR$(2,0)$ is calculated. 
We notice that, in order to speed up the decoding, the values of the LLRs and hard-decisions at the intermediate stages can be stored, since they remain constant during the decoding once calculated.

%% file: numerical_illustrations.tex
In the following, we show the performance of multi-kernel polar codes for the kernels described in Section \ref{sec:mixed_2_3}, i.e., where $T_2$ and $T_3$ kernels are mixed.  

In particular, we show the performance of the multi-kernel polar codes of length $N=72= 2^3 \cdot 3^2$ and $N = 48 = 2^4 \cdot 3$. 
For $N = 72$ there exist 10 possible permutations of the kernels $T_2$ and $T_3$, i.e., 10 different ways to construct the transformation matrix of the multi-kernel polar code. 
For $K=36$ information bits, density evolution analysis suggests to use the transformation matrix $G_{72}=T_3 \otimes T_2 \otimes T_2 \otimes T_2 \otimes T_3$ to build the code. 
For $N = 48$, only 5 permutations of the kernels $T_2$ and $T_3$ can be generated, and the transformation matrix $G_{48}=T_2 \otimes T_2 \otimes T_2 \otimes T_2 \otimes T_3$ has been selected by the density evolution analysis for $K = 24$. 
Both codes are designed for a rate equal to $1/2$ and SNR of 2 $\text{dB}$. 

\begin{figure}
\includegraphics[width=0.48\textwidth]{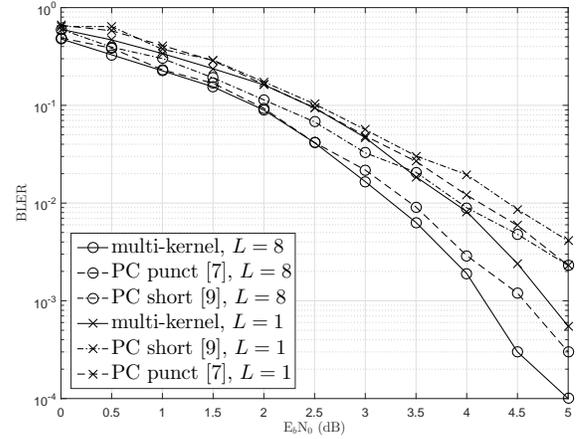}
\caption{Performance of length-72 codes with SCL decoding for rate $1/2$.}
\label{fig:plot_72}
\end{figure}

\begin{figure}
\includegraphics[width=0.48\textwidth]{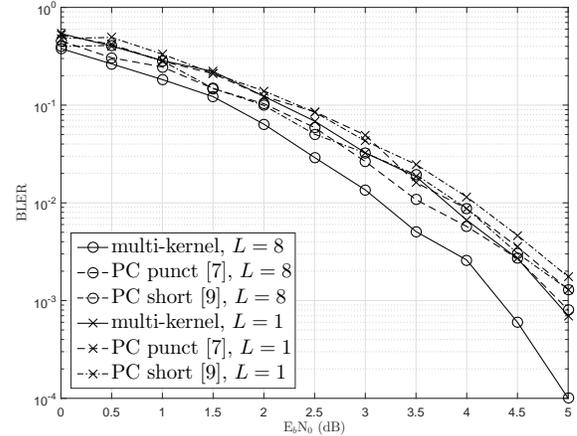}
\caption{Performance of length-48 codes with SCL decoding for rate $1/2$.}
\label{fig:plot_48}
\end{figure}

In Figure~\ref{fig:plot_72} and in Figure~\ref{fig:plot_48} we compare the BLock Error Rate (BLER) performance of the proposed multi-kernel polar code for an additive white Gaussian noise (AWGN) channel against state-of-the-art punctured and shortened polar codes, proposed in \cite{chen_kai_punc} and in \cite{wang_liu}, respectively. 
For $N=72$ a mother polar code of length $N'=128$ is used, while for $N=48$ the mother polar code has length $N'=64$, both designed for the same target SNR of 2 $\text{dB}$. 
In the figures, we show the SC-List decoding performance of the codes for list size $L = 8$ and $L = 1$, the latter corresponding to SC decoding.

We observe that for both cases, our construction significantly outperforms state-of-the-art punctured and shortened polar codes. 
Moreover, multi-kernel polar codes exhibit a smaller decoding complexity compared to punctured/shortened polar codes, due to their reduced code length construction. 
In fact, SC decoding of a punctured/shortened polar code has to be performed over the Tanner graph of its mother polar code, i.e., the complexity of the code depends on the length of the mother code. 
An estimation of the complexity is calculated as the number of LLRs calculated during the decoding, given by the block length multiplied by the number of stages of the graph. 
For multi-kernel polar codes, $N \cdot s$ LLRs are calculated, while for punctured/shortened polar codes $N' \log_2 N'$ LLRs are calculated, with $N' = 2^{\lceil \log_2 N' \rceil}$ since the decoding is performed over the graph of the mother code. 
In particular, for length-$72$ punctured/shortened polar codes, the calculation of 896 LLRS is required, compared to only 360 LLRs for multi-kernel polar codes, which shows a substantial complexity reduction. 

%% file: future_work.tex
In this paper, we proposed a generalized polar code construction based on the polarization of multiple kernels. 
The construction provides numerous practical advantages, as more code lengths can be achieved without puncturing or shortening and only modest puncturing and shortening is required to achieve any arbitrary code length. 
For an example with binary kernels of size 2 and 3, we observed numerically that the error-rate performance of our construction clearly outperforms state-of-the-art constructions using puncturing or shortening methods.